\documentclass{article}
\usepackage{spconf,amsmath,graphicx}
\usepackage{listings}
\usepackage{color}
\usepackage{multirow}
\usepackage{amsmath}
\usepackage{mathtools, nccmath}
\usepackage{booktabs}
\usepackage{textcomp}
\usepackage[ruled,vlined]{algorithm2e}
\usepackage{etoolbox}
\SetArgSty{textnormal}

\DeclarePairedDelimiter{\round}\lfloor\rceil

\definecolor{dkgreen}{rgb}{0,0.6,0}
\definecolor{gray}{rgb}{0.5,0.5,0.5}
\definecolor{mauve}{rgb}{0.58,0,0.82}

\lstdefinestyle{algo1}{
  float=tp,
  floatplacement=tbp,
  abovecaptionskip=-5pt
}

\newcommand\blfootnote[1]{%
  \begingroup
  \renewcommand\thefootnote{}\footnote{#1}%
  \addtocounter{footnote}{-1}%
  \endgroup
}

\definecolor{orange}{rgb}{1,0.5,0}
\definecolor{mdgreen}{rgb}{0.05,0.6,0.05}
\definecolor{mdblue}{rgb}{0,0,0.7}
\definecolor{dkblue}{rgb}{0,0,0.5}
\definecolor{dkgray}{rgb}{0.3,0.3,0.3}
\definecolor{slate}{rgb}{0.25,0.25,0.4}
\definecolor{gray}{rgb}{0.5,0.5,0.5}
\definecolor{ltgray}{rgb}{0.7,0.7,0.7}
\definecolor{purple}{rgb}{0.7,0,1.0}
\definecolor{lavender}{rgb}{0.65,0.55,1.0}

\title{Dynamic Sparsity Neural Networks for Automatic Speech Recognition}
\name{Zhaofeng Wu$^{1}$, Ding Zhao$^2$, Qiao Liang$^2$, Jiahui Yu$^2$, Anmol Gulati$^2$, Ruoming Pang$^2$}
\address{
  $^1$Paul G. Allen School of Computer Science \& Engineering, University of Washington $^2$Google}
\begin{document}
\ninept
\maketitle
\blfootnote{This work was done while Zhaofeng Wu was an intern at Google.}
\begin{abstract}
In automatic speech recognition (ASR), model pruning is a widely adopted technique that reduces model size and latency to deploy neural network models on edge devices with resource constraints. However, multiple models with different sparsity levels usually need to be separately trained and deployed to heterogeneous target hardware with different resource specifications and for applications that have various latency requirements. In this paper, we present Dynamic Sparsity Neural Networks (\emph{DSNN}) that, once trained, can instantly switch to any predefined sparsity configuration at run-time. We demonstrate the effectiveness and flexibility of \emph{DSNN} using experiments on internal production datasets with Google Voice Search data, and show that the performance of a \emph{DSNN} model is on par with that of individually trained single sparsity networks. Our trained \emph{DSNN} model, therefore, can greatly ease the training process and simplify deployment in diverse scenarios with resource constraints.
\end{abstract}
\section{Introduction}

Traditionally, network pruning methods~\cite{han2015learning, zhu2017prune} have been employed to obtain sparse neural network models to support edge devices with limited resources~\cite{shangguan2019optimizing}. However, machine learning production models today often target a variety of consumer hardware capabilities. The wide spectrum of mobile devices alone differ in latency by orders of magnitude for the same machine learning model~\cite{ignatov2018ai}. Systems such as home speakers and cars further increase this disparity. Additionally, different software applications can have distinct latency requirements. For example, the speech recognizer for real-time video conference captioning requires higher synchronicity than one for online video website subtitle generation.

Ideally, different-sized models with varying sparsity levels should be trained to target every single device type. However, this is impractical given the myriad of existing devices. Alternatively, one could train a few sparse models only targeting typical hardware configurations, but it necessitates the maintenance overhead of a device sparsity table. Moreover, even on a single device, resource availability fluctuates as concurrent activities vary. Models with static sparsity levels hence likely lead to sub-optimal resource usage.

To support such diversity of scenarios, we propose Dynamic Sparsity Neural Networks (DSNN). A single trained DSNN model can execute at any predefined sparsity configuration at inference time with no or insignificant loss in accuracy compared to regular individually trained single sparsity networks. DSNN enables dynamic sparsity adjustment according to device capability, resource availability, and application requirements, thereby achieving an optimal accuracy-latency trade-off with minimal memory footprint.

DSNN was inspired by recent work~\cite{zhou2019deconstructing, ramanujan2019whats} which showed that even for untrained random networks, there exist arbitrarily sparse sub-networks that achieve very high quality. Therefore, it is likely that trained networks also simultaneously contain powerful sub-networks at multiple sparsity levels.

Methodologically, DSNN builds upon slimmable neural networks (SNN)~\cite{yu2018slimmable, yu2019universally} that similarly tackle model deployment across heterogeneous devices. However, SNN was only designed for convolutional neural networks, restricting their applicability to many domains and tasks. We demonstrate in \S\ref{sec:snn-vs-dsnn} that a naive generalization of SNN to automatic speech recognition (ASR) models shows poor performance.

DSNN, on the other hand, is a sparsity-based extension of SNN that is applicable to any weight-based neural network. Since modern specialized hardware allows such sparse models to have comparable speedup to SNN that prunes entire convolutional channels~\cite{shangguan2019optimizing}, this generalization comes at little inference time cost. In this paper, we choose to focus on the task of ASR due to an increasing demand for on-device ASR~\cite{he2019streaming, shangguan2019optimizing}. We show that a single DSNN model generally matches the quality of individually trained single sparsity networks across multiple sparsity configurations (\S\ref{sec:baseline-comparison}). DSNN hence contributes to practical machine learning systems through its ability to dynamically adjust to multiple hardware types with different resource and energy constraints. This greatly reduces both the training overhead and the management complexity of deployment processes. 

\section{Dynamic Sparsity Neural Networks}
\label{sec:dsnn}

In this section, we first briefly introduce regular sparse neural networks. We then provide a formulation of dynamic sparsity neural networks (DSNN) and justify it using previous studies. Next, we introduce the DSNN training algorithm. Finally we sketch several key distinctions with slimmable neural networks, our methodological precursor.

\subsection{Sparse Neural Networks} \label{sec:sparse-nn}

Model pruning removes connections in a neural network, yielding a lighter yet static inference time model. It usually follows a three-step procedure~\cite{liu2018rethinking}: (1) initialize from a pretrained over-parameterized full model; (2) remove certain connections based on a criterion; (3) fine-tune the remaining weights. Often the pruning happens gradually, alternating between (2) and (3) until a single given target sparsity level $S \in [0, 1)$ is reached~\cite{zhu2017prune}. In this work, we employ gradient-based pruning, i.e., using the $L_1$ norm of (weight $\times$ gradient) as the pruning criterion~\cite{lee2018snip}. At each step, for each weight $W$, $S|W|$ elements with the smallest such norms are zeroed out by applying a binary mask $M$ over $W$. We allow $M$ to update at each iteration, enabling pruned weights to be recovered if at a later step its norm is greater than that of some survived weight~\cite{guo2016dynamic, he2018soft}.%

\subsection{Model Formulation} \label{sec:formulation}

\begin{figure}[t]
  \centering
  \includegraphics[width=0.97\linewidth]{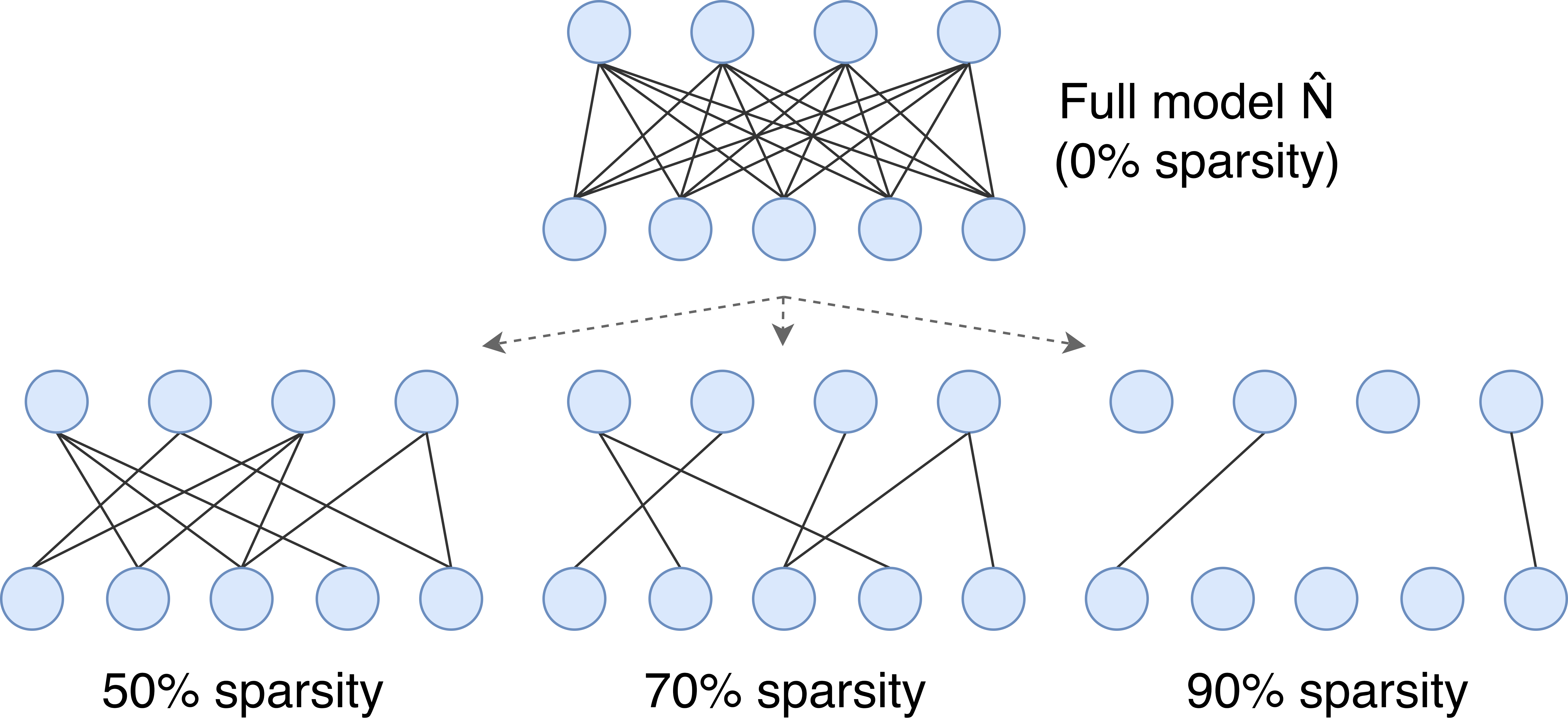}
  \caption{Illustration of dynamic sparsity neural networks. Once trained, the super-network can execute at any given sparsity configuration, or keep its original full capacity, at inference time.}
  \label{fig:dsnn}
\end{figure}

Let a sparsity configuration $C$ represent a set of sparsity levels $S_W$ for every weight $W$ in a network.
DSNN takes as input a list of such sparsity configurations $[C_0, \cdots, C_L]$, where we stipulate $C_0$ to always be the full network, i.e., $\forall S \in C_0, S = 0\%$.
DSNN aims to train a super-network $\hat{N}$ that, given \emph{any} sparsity configuration $C_i$ in the list, can induce a sub-network $\hat{N}_{C_i}$ with a subset of connections, without further fine-tuning. The sub-network $\hat{N}_{C_i}$ should have similar quality to an individually trained single sparsity model $\tilde{N}_{C_i}$, obtained through traditional pruning algorithms from $\hat{N}$ (\S\ref{sec:sparse-nn}), at the same sparsity configuration $C_i$. With this super-network, we can dynamically switch to any sub-network with different sparsity configurations when deployed, optimizing for hardware capacities and application latency constraints. We illustrate this process in Figure~\ref{fig:dsnn}.%

Empirical evidence suggests the likely existence of $\hat{N}$. \cite{zhou2019deconstructing} and \cite{ramanujan2019whats} showed that an \textit{untrained} random model simultaneously contains sub-networks that perform well. At specific sparsity levels, these models perform as well as individually trained dense models. Slimmable neural networks~\cite{yu2018slimmable, yu2019universally} showed that for ImageNet~\cite{deng2009imagenet} classification, a trained convolutional super-network can have structured sub-networks with similar performance to individually trained networks. BigNAS~\cite{yu2020bignas} trained a single set of shared weights on ImageNet which are used to obtain child models via a simple coarse-to-fine architecture selection heuristic. All these works suggest a possible super-network that encompasses multiple high quality sub-networks and encourage us to explore DSNN, a general sparsity-based super-network.

\subsection{Approach} \label{sec:approach}

\makeatletter
\patchcmd{\@algocf@start}%
  {-1.5em}%
  {0pt}%
  {}{}%
\makeatother
\begin{algorithm}[t]
\SetKwInput{KwInput}{Input}
\SetKw{KwInit}{Initialize}
\SetKw{KwIn}{in}
\DontPrintSemicolon
\KwInput{sparsity configurations C$_1$ ... C$_L$, num epochs T, num progressive freezing epochs T', mask update frequency F}
\KwInit pretrained full model N\;
\KwInit binary mask M[W] for each weight W in N with 1\;
\KwInit grad[W] for each weight W in N with 0\;

\For{epoch $\leftarrow$ 0, ..., T -- 1} {
  $y'$ = forward(N) \tcp*{full model}
  curr\_grad = backward(loss($y'$, $y$))\;
  \For(\tcp*[f]{$\uparrow$ sparsity}){C $\leftarrow$ sorted(C$_1$ ... C$_L$)}{
    \If{epoch \% F == 0} {
      \For{W \KwIn N} {
        M[W] = get\_mask(W, grad[W], C[W])\;
      }
    }
    $\hat{y}$ = forward(N $\circ$ M)\;
    curr\_grad += backward(loss($\hat{y}$, $y'$))\;
  }
  grad = curr\_grad\;
  Update all weights with optimizer using grad\;
}
\For{W \KwIn N} {
  Set M[W] to 1 iff get\_mask(W, grad[W], C[W]) is 1 for any C in C$_1$ ... C$_L$\;
}
\For(\tcp*[f]{prog. freezing}){epoch $\leftarrow$ T, ..., T + T' -- 1} {
  $y'$ = forward(stop\_gradient(N $\circ$ M) + N $\circ$ !M)\;
  grad = backward(loss($y'$, $y$))\;
  Update all weights with optimizer using grad\;
}
\caption{Dynamic sparsity neural networks training. The get\_mask(weight, gradient, sparsity) function, described in \S\ref{sec:sparse-nn}, returns a binary mask which is 0 iff the corresponding element is among the sparsity$\times\vert$weight$\vert$ elements with the smallest weight$\times$gradient norm. The stop\_gradient(variable) function is equivalent to zeroing out the gradient of the loss w.r.t. this variable during optimization.}
\label{lst:DSNN-algo}
\end{algorithm}

With the likely existence of a super-network $\hat{N}$, the question becomes how to efficiently find it. For a single model to execute at multiple sparsity configurations at inference time, we jointly train the same network with these configurations. In each training epoch, we alternate between these configurations, as well as the full model $C_0$, for weight masking and execute forward and backward propagation. As we allow the mask to update between iterations, this alternation also has a regularization effect, though it is not our ultimate purpose, by preventing weight co-adaptation, similar to DropConnect~\cite{pmlr-v28-wan13}.

We discover, however, that this alternation during training causes convergence instability, for example suffering from frequent gradient explosion. We employ two forms of lazy update to alleviate this issue. First, during each epoch, we accumulate gradients across all sparsity configurations and only update the parameters at the end of the epoch when all configurations are traversed. Second, we do not update the mask in every epoch, but train the same weights across $F$ epochs regardless of whether or not their updated norm makes them ineligible to be retained.

We take inspiration from regular sparse model training and pretrain the full model before pruning begins~\cite{luo2017thinet, carreira2018learning}. This gives a high performance first-step model as the initialization for sparse models to prune from. For DSNN, although we always include the the full model $C_0$ in each epoch, empirically we find pretraining to still be crucial for the DSNN quality.

The simultaneous presence of networks of multiple sizes naturally enables in-place distillation~\cite{yu2019universally} by transferring the knowledge of large networks to smaller ones. We use a similar distillation method as~\cite{Panchapagesan20} except DSNN allows the distillation process to happen in one shot by computing the distillation loss for smaller networks using the full network's output probabilities.

In a production setting, the full model, operating without resource constraints, is usually the most valued and hence has the least tolerance for quality degradation. We, therefore, further fine-tune the full model after DSNN convergence. During the backward propagation, we ignore gradient contributed by the weights in smaller networks (so that those networks remain the same quality) and update only the remaining weights. This yields a slight quality gain for the full model (\S\ref{sec:ablations}). We call this scheme ``progressive freezing.''

We sketch the entire DSNN training procedure in Algorithm~\ref{lst:DSNN-algo}.

\subsection{Comparison with Slimmable Neural Networks}

Our model is similar to slimmable neural networks (SNN)~\cite{yu2018slimmable, yu2019universally}, both allowing dynamic inference graphs, albeit with several key distinctions. Most importantly, SNN shrinks models by truncating convolutional channels while DSNN obtains smaller model variants using model pruning. This allows DSNN to be easily applied to more domains and tasks. We may consider a simple generalization of SNN that prunes whole nodes in a network instead of convolutional channels. In contrast, DSNN uses an edge-pruning approach, the common practice for model pruning. This restricts SNN to always use fully connected sub-networks. On the other hand, the lack of a predefined network structure in DSNN allows greater modeling flexibility. DSNN's sparse structure allows it to preserve the high dimensionality of input and output spaces, although the mapping from input to output is low-dimensional. Therefore, this generalized SNN is a special case of DSNN whose sparse patterns are skewed with all connections to the last channels masked as zeros. Below we use ``SNN'' to refer to this generalization.\footnote{We experimented with the "sandwich rule" technique~\cite{yu2019universally} that enables inference with arbitrary sparsity levels beyond predefined levels. However, we observed slight performance degradation and hence did not adapt it. In production, it is usually unnecessary to generalize beyond predefined sparsity levels with reasonable downstream scenario coverage.}

\section{Experimental Setup}

In this section, we first describe the task and datasets with which we experiment. We then introduce the architecture backbone as well as the pruning settings that operate on it. %

\subsection{Task and Dataset}

While our approach is widely applicable to all weight-based neural networks, we focus on automatic speech recognition (ASR) due to an increased interest in on-device ASR~\cite{he2019streaming, shangguan2019optimizing}. We use the same in-house production training dataset as \cite{Zhao19} which consists of 35 million English utterances (\texttildelow 27,500 hours), representative of Google's voice search traffic, that are anonymized and hand-transcribed. These utterances are then artificially corrupted with noise and reverberation~\cite{Chanwoo17}. We refer readers to \cite{Zhao19} for more details. We use two separate internal test sets to evaluate the model quality. One contains around 15,000 English utterances of Google's voice search traffic (\textbf{VS}) and the other consists of 9,000 noisy farfield utterances where the sound source is far from the microphone (\textbf{Farfield}). We use word error rate (WER) to measure model quality.

\subsection{Model Architecture and Settings}

We follow the model architecture of \cite{he2019streaming}. It has the Recurrent Neural Network Transducer (RNN-T) model~\cite{Graves12} as the backbone. The encoder contains a time reduction layer~\cite{Soltau2017} followed by 8 2,000-dim LSTM layers and a 600-dim projection layer. The decoder contains 2 2,000-dim LSTM layers with a 600-dim projection layer in each LSTM layer. The encoder and decoder are fed into a 600-dim joint-network and then to a 4,096-dim softmax layer. We use a constant learning rate of 1e-3, after warm-up, with the Adam optimizer~\cite{kingma2014adam}. 
We maintain exponential moving averages of the trained parameters and use the averaged parameters during evaluation. 
We refer readers to \cite{he2019streaming} for more details.
The models are trained in Tensorflow~\cite{AbadiAgarwalBarhamEtAl15} on $8 \times 8$ Tensor Processing Unit (TPU) slices with a batch size of 4,096.

\subsection{Model Pruning Settings}
\label{sec:model-pruning}

\begin{table}[t]
  \caption{Target sparsity configurations. The ``Average'' columns represents an average global sparsity level across all weights.}
  \label{tab:sparsity-configs}
  \centering
  \begin{tabular}{ c c c c c }
    \toprule
    & \multicolumn{3}{c}{\textbf{Sparsity}} & \\
    \cmidrule(lr){2-4}
    \textbf{Type} & \textbf{LSTM} & \textbf{FC} & \textbf{Average} & \textbf{\# Parameters}  \\
    \midrule
    Large & \phantom{0}0\% & \phantom{0}0\% & \phantom{0}0\% & 122.2M \\
    Medium & 70\% & \phantom{0}0\% & 68\% & \phantom{0}39.6M \\
    Small & 90\% & 50\% & 88\% & \phantom{0}14.6M \\
    \bottomrule
  \end{tabular}
  
\end{table}

We prune the model to different sizes for different use cases: a large model can be used for servers and high-end devices which do not require any pruning; on the other hand, lower-end phones can benefit from a sparser model.
We prune all 2D matrices in LSTM and fully-connected (FC) layers, which contain 98.7\% of all model parameters. We allow different sparsity levels for LSTM and FC layers and carefully select three sparsity configurations that best fit downstream resource constraints, including one that corresponds to the full model $C_0$ (``Large'').
Table~\ref{tab:sparsity-configs} outlines these sparsity configurations.

As baselines, we train \emph{separate} single sparsity models under these three sparsity configurations, as well as a DSNN with all configurations.
For all models, we first train a zero-sparsity network until convergence (\texttildelow600k steps).
We maintain exponential moving averages of all model parameters during pretraining for the initialization of the pruning stage. 
All models use a cubic schedule to gradually increase the training sparsity from 0 to the target sparsity during the first 100k steps.

To efficiently leverage hardware resources, we employ block pruning~\cite{narang2017block} with block size $16 \times 1$. Instead of the smallest elements, we zero out the smallest $16 \times 1$ blocks in $W$.
After pruning, we convert sparse matrices to a dense representation in TensorFlow-Lite and use custom operations to speed up sparse matrix multiplications which allows proportional speedup to increasing sparsity~\cite{shangguan2019optimizing}.

\section{Results and Discussion}

In this section, we present experimental results that compare DSNN to both individually trained single sparsity networks and SNN as well as ablation results.

\begin{table}[t]
  \caption{Word error rate (WER) for single sparsity networks (Single), DSNN, and SNN on the two test sets. Lower is better. Both DSNN and SNN use one model across all three configurations. The ``Sparsity'' column represents the average global sparsity.}
  \label{tab:main-results}
  \centering
  \begin{tabular}{ c @{\hspace{4pt}} c @{\hspace{7pt}} c | c c c c }
    \toprule
    \textbf{Type} & \textbf{Sparsity} & \textbf{Model} & \textbf{VS} & \textbf{Farfield} \\
    \midrule
    \multirow{3}{*}{Large} & \multirow{3}{*}{\phantom{0}0\%} & Single$_\text{L}$ & \phantom{0}5.9 & \phantom{0}4.2  \\
    && DSNN & \phantom{0}6.0 & \phantom{0}4.2 \\
    && SNN & \phantom{0}6.0 & \phantom{0}4.2 \\
    \midrule
    \multirow{3}{*}{Medium} & \multirow{3}{*}{68\%} & Single$_\text{M}$ & \phantom{0}6.9 & \phantom{0}5.0 \\
    && DSNN & \phantom{0}7.2 & \phantom{0}5.4  \\
    && SNN & \phantom{0}9.8 & \phantom{0}7.5 \\
    \midrule
    \multirow{3}{*}{Small} & \multirow{3}{*}{88\%} & Single$_\text{S}$ & \phantom{0}9.5 & \phantom{0}7.0 \\
    && DSNN & \phantom{0}9.9 & \phantom{0}7.6 \\
    && SNN  & 34.3 & 27.5 \\
    \bottomrule
  \end{tabular}
  
\end{table}

\subsection{Comparison with Single Sparsity Networks} \label{sec:baseline-comparison}

As argued in \S\ref{sec:formulation}, DSNN should match individually trained single sparsity networks in quality. 
We, therefore, compare DSNN with single sparsity networks and show the results in Table~\ref{tab:main-results}. The dynamic sparsity model generally matches the quality of single sparsity networks, even up to 88\% global sparsity. In particular, for the large (i.e. full) model whose quality is usually prioritized in production, DSNN and a regularly trained model perform almost identically.

\subsection{Comparison with Slimmable Neural Networks} \label{sec:snn-vs-dsnn}

Slimmable neural networks (SNN) only vary the number of channels in convolutional networks. For each target width, the last channels (with the largest indices) are removed. Despite not directly applicable to arbitrary weight matrices, we experiment with a simple generalization of SNN as follows.

Given a target sparsity level $S \in [0, 1)$ and a weight matrix $W \in \mathcal{R}^{N_1 \times N_2 \times \cdots \times N_D}$, we apply a binary mask $M$ on $W$. For each dimension $d \in [1, D]$, we select a threshold $T_d$ with

\begin{equation}
T_d = \round*{N_d * (1 - S) ^ \frac{1}{D}}
\end{equation}
where $\round*{\cdot}$ rounds to the nearest integer. We then generate $M$ by

\begin{equation}
M[i_1, ..., i_D] = \begin{cases}
    1 & \text{if } \forall d \in [1, D], i_d \le T_d \\
    0 & \text{otherwise}
\end{cases}
\end{equation}
Finally, we prune the weight $W = M \circ W$.

Intuitively, we truncate the last rows of $W$ in each dimension by an equal fraction trying to make the resulting matrix have a sparsity close to the target sparsity.
This is analogous to removing the last convolutional channels.

We compare SNN and DSNN quality in Table~\ref{tab:main-results}. We see that DSNN significantly outperforms SNN. We hypothesize that this is due to the better flexibility of DSNN's edge pruning approach compared to SNN's node pruning approach, as well as DSNN's more informed pruning choices rather than always removing the last rows.

\subsection{Ablations}
\label{sec:ablations}

\begin{table}[t]
\small
  \caption{Word error rate (WER) for incrementally adding gradient accumulation, in-place distillation, and progressive freezing on top of a naive DSNN implementation on the two test sets. Lower is better. The ``Sparsity'' column represents the average global sparsity.}
  \label{tab:ablations}
  \centering
  \begin{tabular}{@{\hspace{5pt}} c @{\hspace{4pt}} c @{\hspace{7pt}} l | c @{\hspace{9pt}} c @{\hspace{5pt}}}
    \toprule
    \textbf{Type} & \textbf{Sparsity} & \textbf{Model} & \textbf{VS} & \textbf{Farfield} \\
    \midrule
    \multirow{4}{*}{Large} & \multirow{4}{*}{\phantom{0}0\%} & Baseline DSNN & 6.4 & 4.5  \\
    && \ \ + Lazy Update & 6.2 & 4.4 \\
    && \ \ \ \ + In-Place Distillation & 6.1 & 4.4 \\
    && \ \ \ \ \ \ + Progressive Freezing & \textbf{6.0} & \textbf{4.2} \\

    \midrule
    \multirow{4}{*}{Medium} & \multirow{4}{*}{68\%} & Baseline DSNN & 7.9 & 5.6  \\
    && \ \ + Lazy Update & 7.8 & 5.7 \\
    && \ \ \ \ + In-Place Distillation & \textbf{7.2} & \textbf{5.4} \\
    && \ \ \ \ \ \ + Progressive Freezing & \textbf{7.2} & \textbf{5.4} \\
    
    \midrule
    \multirow{4}{*}{Small} & \multirow{4}{*}{88\%} & Baseline DSNN & 11.9 & 8.7  \\
    && \ \ + Lazy Update & 10.5 & 8.2 \\
    && \ \ \ \ + In-Place Distillation & \textbf{9.9} & \textbf{7.6} \\
    && \ \ \ \ \ \ + Progressive Freezing & \textbf{9.9} & \textbf{7.6} \\

    \bottomrule
  \end{tabular}
\end{table}

We conduct ablation experiments for the training techniques described in \S\ref{sec:approach} and report the results in Table~\ref{tab:ablations}. Lazy update, including gradient accumulation and lazy mask update, helps stabilize training for the large and medium models but without much quality differences. It, however, significantly improves the small model quality. We hypothesize that a stable training procedure could help find a better local minimum for very sparse models. In-place distillation uniformly improves the model quality across the board, especially for the medium and small models which are the configurations that this technique targets. Continuing training part of the large model after the smaller ones converge also slightly improves its performance without sacrificing the quality of smaller models. This helps further close the gap between DSNN and the single sparsity baseline.

\section{Related Work}

Over-parameterization is a commonly addressed issue of neural networks~\cite{zhang2016understanding, arpit2017closer}. To deal with this issue, model pruning methods have been developed to remove unimportant connections in weight matrices of neural network models~\cite{NIPS1989_250,hassibi1993second,han2015learning}.
The resulting pruned models contain only sparse structures, allowing them to run efficiently at inference time while maintaining performance. Many studies have demonstrated the empirical strength of such sparse networks~\cite{han2015learning, shangguan2019optimizing} and examined their theoretical properties~\cite{frankle2018lottery, frankle2019stabilizing, zhou2019deconstructing, ramanujan2019whats}.

Recently,~\cite{zhou2019deconstructing} and~\cite{ramanujan2019whats} showed that untrained random networks contain sub-networks at arbitrary sparsity levels that perform well without training. The best of these sub-networks, usually at around 50\% sparsity, can perform as well as the full (i.e. zero-sparsity) model in specific datasets. Our work also tries to find a single network containing multiple high-quality sub-networks, but we allow model training while requiring these sub-networks to match the quality of individually-trained single sparsity networks.

Dynamic neural networks are a family of models that optimize the run-time accuracy and efficiency trade-off using dynamic inference graphs~\cite{liu2017dynamic, huang2017multiscale, yu2018slimmable, yu2019universally}. These models usually allow selective execution which is desirable when the target inference platforms vary in their resource constraints. To our knowledge, our proposed DSNN is the first of such models that achieves such optimization using sparse networks, a more general approach.

\section{Conclusion and Future Work}

We presented a training scheme that allows one single trained model to optimally switch its sparsity level at inference time. Given that its performance is on par with individually trained single sparsity networks, such a model can simultaneously support a variety of devices with different hardware capabilities and applications with diverse latency requirements. Future work can attempt to close the marginal gap between DSNN and single sparsity networks, especially at high sparsity levels.

In this work, we only considered models with all parameters of the same layer type pruned by the same fraction. However, components of a machine learning model are sometimes not equally important and setting different sparsity levels for different weights may yield a higher quality model~\cite{see2016compression}. As each weight matrix is independently pruned in the DSNN training algorithm, DSNN is able to approximate the performance of individually trained networks with arbitrary sparsity configurations across weights. Combined with a greedy search algorithm, DSNN can be used to search for an optimal per-weight sparsity configuration, similar to~\cite{yu2019network}. This can be an interesting future exploration.

\bibliographystyle{IEEEbib}
\bibliography{mybib}

\begin{thebibliography}{10}

\bibitem{han2015learning}
S.~Han, J.~Pool, J.~Tran, and W.~Dally,
\newblock ``Learning both weights and connections for efficient neural
  network,''
\newblock in {\em Proc. of NIPS}, 2015.

\bibitem{zhu2017prune}
M.~H. Zhu and S.~Gupta,
\newblock ``To prune, or not to prune: Exploring the efficacy of pruning for
  model compression,'' 2018.

\bibitem{shangguan2019optimizing}
Y.~Shangguan, J.~Li, L.~Qiao, R.~Alvarez, and I.~McGraw,
\newblock ``Optimizing speech recognition for the edge,''
\newblock {\em arXiv preprint arXiv:1909.12408}, 2019.

\bibitem{ignatov2018ai}
A.~Ignatov, R.~Timofte, W.~Chou, K.~Wang, M.~Wu, T.~Hartley, and L.~Van~Gool,
\newblock ``{AI} benchmark: Running deep neural networks on android
  smartphones,''
\newblock in {\em Proc. of ECCV}, 2018.

\bibitem{zhou2019deconstructing}
H.~Zhou, J.~Lan, R.~Liu, and J.~Yosinski,
\newblock ``Deconstructing lottery tickets: Zeros, signs, and the supermask,''
\newblock in {\em Proc. of NeurIPS}, 2019.

\bibitem{ramanujan2019whats}
V.~Ramanujan, M.~Wortsman, A.~Kembhavi, A.~Farhadi, and M.~Rastegari,
\newblock ``What's hidden in a randomly weighted neural network?,''
\newblock in {\em Proc. of CVPR}, 2020.

\bibitem{yu2018slimmable}
J.~Yu, L.~Yang, N.~Xu, J.~Yang, and T.~Huang,
\newblock ``Slimmable neural networks,''
\newblock in {\em Proc. of ICLR}, 2019.

\bibitem{yu2019universally}
J.~Yu and T.~Huang,
\newblock ``Universally slimmable networks and improved training techniques,''
\newblock in {\em Proc. of ICCV}, 2019.

\bibitem{he2019streaming}
Y.~He, T.~N. Sainath, R.~Prabhavalkar, I.~McGraw, R.~Alvarez, D.~Zhao,
  D.~Rybach, A.~Kannan, Y.~Wu, R.~Pang, et~al.,
\newblock ``Streaming end-to-end speech recognition for mobile devices,''
\newblock in {\em Proc. of ICASSP}, 2019.

\bibitem{liu2018rethinking}
Z.~Liu, M.~Sun, T.~Zhou, G.~Huang, and T.~Darrell,
\newblock ``Rethinking the value of network pruning,''
\newblock in {\em Proc. of ICLR}, 2019.

\bibitem{lee2018snip}
N.~Lee, T.~Ajanthan, and P.~Torr,
\newblock ``{SNIP}: Single-shot network pruning based on connection
  sensitivity,''
\newblock in {\em Proc. of ICLR}, 2019.

\bibitem{guo2016dynamic}
Y.~Guo, A.~Yao, and Y.~Chen,
\newblock ``Dynamic network surgery for efficient {DNN}s,''
\newblock in {\em Proc. of NIPS}, 2016.

\bibitem{he2018soft}
Y.~He, G.~Kang, X.~Dong, Y.~Fu, and Y.~Yang,
\newblock ``Soft filter pruning for accelerating deep convolutional neural
  networks,''
\newblock in {\em Proc. of IJCAI}, 2018.

\bibitem{deng2009imagenet}
J.~Deng, W.~Dong, R.~Socher, L.~Li, K.~Li, and F.~Li,
\newblock ``Imagenet: A large-scale hierarchical image database,''
\newblock in {\em Proc. of CVPR}, 2009.

\bibitem{yu2020bignas}
J.~Yu, P.~Jin, H.~Liu, G.~Bender, P.~Kindermans, M.~Tan, T.~Huang, X.~Song,
  R.~Pang, and Q.~Le,
\newblock ``{BigNAS}: Scaling up neural architecture search with big
  single-stage models,''
\newblock {\em arXiv preprint arXiv:2003.11142}, 2020.

\bibitem{pmlr-v28-wan13}
L.~Wan, M.~Zeiler, S.~Zhang, Y.~LeCun, and R.~Fergus,
\newblock ``Regularization of neural networks using dropconnect,''
\newblock in {\em Proc. of ICML}, 2013.

\bibitem{luo2017thinet}
J.~Luo, J.~Wu, and W.~Lin,
\newblock ``{ThiNet}: A filter level pruning method for deep neural network
  compression,''
\newblock in {\em Proc. of ICCV}, 2017.

\bibitem{carreira2018learning}
M.~A. Carreira-Perpin{\'a}n and Y.~Idelbayev,
\newblock ``{``Learning-compression''} algorithms for neural net pruning,''
\newblock in {\em Proc. of CVPR}, 2018.

\bibitem{Panchapagesan20}
S.~Panchapagesan, D.~S. Park, C.~Chiu, Y.~Shangguan, Q.~Liang, and
  A.~Gruenstein,
\newblock ``Efficient knowledge distillation for rnn-transducer models,'' 2020.

\bibitem{Zhao19}
D.~Zhao, T.~N. Sainath, D.~Rybach, P.~Rondon, D.~Bhatia, B.~Li, and R.~Pang,
\newblock ``Shallow-fusion end-to-end contextual biasing,''
\newblock in {\em Proc. Interspeech}, 2017.

\bibitem{Chanwoo17}
C.~Kim, A.~Misra, K.~Chin, T.~Hughes, A.~Narayanan, T.~N. Sainath, and
  M.~Bacchiani,
\newblock ``Generated of large-scale simulated utterances in virtual rooms to
  train deep-neural networks for far-field speech recognition in {Google
  Home},''
\newblock in {\em Proc. Interspeech}, 2017.

\bibitem{Graves12}
A.~Graves,
\newblock ``Sequence transduction with recurrent neural networks,''
\newblock in {\em Proc. of ICML}, 2012.

\bibitem{Soltau2017}
H.~Soltau, H.~Liao, and H.~Sak,
\newblock ``Reducing the computational complexity for whole word models,''
\newblock in {\em Proc. of ASRU}, 2017.

\bibitem{kingma2014adam}
D.~P. Kingma and J.~Ba,
\newblock ``Adam: A method for stochastic optimization,''
\newblock in {\em Proc. of ICLR}, 2015.

\bibitem{AbadiAgarwalBarhamEtAl15}
M.~Abadi, A.~Agarwal, P.~Barham, E.~Brevdo, Z.~Chen, C.~Citro, G.~S. Corrado,
  A.~Davis, J.~Dean, M.~Devin, et~al.,
\newblock ``Tensorflow: Large-scale machine learning on heterogeneous
  distributed systems,'' 2015.

\bibitem{narang2017block}
S.~Narang, E.~Undersander, and G.~Diamos,
\newblock ``Block-sparse recurrent neural networks,''
\newblock {\em arXiv:1711.02782}, 2017.

\bibitem{zhang2016understanding}
C.~Zhang, S.~Bengio, M.~Hardt, B.~Recht, and O.~Vinyals,
\newblock ``Understanding deep learning requires rethinking generalization,''
  2017.

\bibitem{arpit2017closer}
D.~Arpit, S.~Jastrzebski, N.~Ballas, D.~Krueger, E.~Bengio, M.~S. Kanwal,
  T.~Maharaj, A.~Fischer, A.~Courville, Y.~Bengio, and S.~Lacoste-Julien,
\newblock ``A closer look at memorization in deep networks,''
\newblock in {\em Proc. of ICML}, 2017.

\bibitem{NIPS1989_250}
Y.~LeCun, J.~S. Denker, and S.~A. Solla,
\newblock ``Optimal brain damage,''
\newblock in {\em Proc. of NIPS}, 1990.

\bibitem{hassibi1993second}
B.~Hassibi and D.~G. Stork,
\newblock ``Second order derivatives for network pruning: Optimal brain
  surgeon,''
\newblock in {\em Proc. of NIPS}, 1993.

\bibitem{frankle2018lottery}
J.~Frankle and M.~Carbin,
\newblock ``The lottery ticket hypothesis: Finding sparse, trainable neural
  networks,''
\newblock in {\em Proc. of ICLR}, 2019.

\bibitem{frankle2019stabilizing}
J.~Frankle, G.~K. Dziugaite, D.~M. Roy, and M.~Carbin,
\newblock ``Stabilizing the lottery ticket hypothesis,'' 2019.

\bibitem{liu2017dynamic}
L.~Liu and J.~Deng,
\newblock ``Dynamic deep neural networks: Optimizing accuracy-efficiency
  trade-offs by selective execution,''
\newblock in {\em Proc. of AAAI}, 2018.

\bibitem{huang2017multiscale}
G.~Huang, D.~Chen, T.~Li, F.~Wu, L.~V. Maaten, and K.~Weinberger,
\newblock ``Multi-scale dense networks for resource efficient image
  classification,''
\newblock in {\em Proc. of ICLR}, 2018.

\bibitem{see2016compression}
A.~See, M.~Luong, and C.~D. Manning,
\newblock ``Compression of neural machine translation models via pruning,''
\newblock in {\em Proc. of CoNLL}, Aug. 2016.

\bibitem{yu2019network}
J.~Yu and T.~Huang,
\newblock ``Network slimming by slimmable networks: Towards one-shot
  architecture search for channel numbers,''
\newblock {\em arXiv preprint arXiv:1903.11728}, 2019.

\end{thebibliography}

\end{document}